\documentclass[preprint,showpacs,preprintnumbers,amsmath,amssymb]{revtex4}
\usepackage{graphicx}
\usepackage{dcolumn}
\usepackage{bm}

\begin{document}

\title{Fragmentation of $^7$Li Relativistic Nuclei on a Proton into the $^3$H+$^4$He Channel}

\author{N.~G.~Peresadko}
   \affiliation{Lebedev Institute of Physics, Russian Academy of Sciences, Leninskii pr. 53, Moscow, 119991 Russia
Received February 10, 2010; in final form, April 19, 2010} 
\author{Yu.~A.~Alexandrov}
   \email{alexandrov@sci.lebedev.ru}
   \affiliation{Lebedev Institute of Physics, Russian Academy of Sciences, Leninskii pr. 53, Moscow, 119991 Russia
Received February 10, 2010; in final form, April 19, 2010} 
 \author{S.~G.~Gerasimov}
   \affiliation{Lebedev Institute of Physics, Russian Academy of Sciences, Leninskii pr. 53, Moscow, 119991 Russia
Received February 10, 2010; in final form, April 19, 2010} 
\author{V.~A.~Dronov}
   \affiliation{Lebedev Institute of Physics, Russian Academy of Sciences, Leninskii pr. 53, Moscow, 119991 Russia
Received February 10, 2010; in final form, April 19, 2010}
\author{V.~G.~Larionova}
   \affiliation{Lebedev Institute of Physics, Russian Academy of Sciences, Leninskii pr. 53, Moscow, 119991 Russia
Received February 10, 2010; in final form, April 19, 2010}
\author{A.~V.~Pisetskaya}
   \affiliation{Lebedev Institute of Physics, Russian Academy of Sciences, Leninskii pr. 53, Moscow, 119991 Russia
Received February 10, 2010; in final form, April 19, 2010}   
\author{A.~I.~Malakhov}
   \affiliation{Lebedev Institute of Physics, Russian Academy of Sciences, Leninskii pr. 53, Moscow, 119991 Russia
Received February 10, 2010; in final form, April 19, 2010} 
\author{E.~I.~Tamm}
   \affiliation{Lebedev Institute of Physics, Russian Academy of Sciences, Leninskii pr. 53, Moscow, 119991 Russia
Received February 10, 2010; in final form, April 19, 2010}
\author{V.~N.~Fetisov}
   \affiliation{Lebedev Institute of Physics, Russian Academy of Sciences, Leninskii pr. 53, Moscow, 119991 Russia
Received February 10, 2010; in final form, April 19, 2010} 
\author{S.~P.~Kharlamov}
   \affiliation{Lebedev Institute of Physics, Russian Academy of Sciences, Leninskii pr. 53, Moscow, 119991 Russia
Received February 10, 2010; in final form, April 19, 2010}   
\author{L.~N.~Shesterkina}
   \affiliation{Lebedev Institute of Physics, Russian Academy of Sciences, Leninskii pr. 53, Moscow, 119991 Russia
Received February 10, 2010; in final form, April 19, 2010}

\begin{abstract}
\indent 
In a track nuclear photoemulsion exposed to a beamof $^7$Li nuclei accelerated to a momentum of 3~GeV/$c$ per nucleon at the synchrophasotron of the Joint Institute for Nuclear Research (JINR, Dubna), 13 events in which $^7$Li nuclei interacting with protons break up into $^3$H and $^4$He fragments were detected among 3730 inelastic-interaction events. For this fragmentation channel, the cross section was found to be $8\pm2$~mb. The average value of the fragment total transverse momentum was $214\pm5$~MeV/$c$. This value exceedsmarkedly the average value of the transverse-momentumtransfer in the coherent dissociation of $^7$Li nuclei on track-emulsion nuclei ($166\pm5$~MeV/$c$). The recoil-proton transverse momentum was on average 98\% of the total proton momentum. The longitudinal-momentum distribution of protons was characterized by a variance of 16~MeV/$c$ and a mean value of $37\pm2$~MeV/$c$.\par
\indent \par
\indent DOI: 10.1134/S1063778810110177\par
\end{abstract}

\maketitle
\section{\label{sec:level1}INTRODUCTION}

\indent Inelastic interactions of nucleons or nuclei with nuclei at high energies (corresponding to momenta of a few GeV/$c$ units per nucleon) for the case where one reaction channel is singled out have been studied much less adequately than elastic scattering. The theory of inelastic reactions at these energies is based on various versions of the Coulombmechanism of the excitation of nuclei \cite{Bertulani} and on the Glauber-Sitenko diffraction theory of multiple collisions \cite{Sitenko}. In practical applications, all of these approaches employ various approximations and therefore call for an experimental verification. In the overwhelming majority of studies devoted to considering inelastic collisions, the deuteron, which is the simplest nucleus, is used as a probing nucleus incident to the target. However, the presence of a neutron in the final state and the uncertainty in the states of the target nucleus after a collision event complicate experimental verification of the aforementioned theoretical approaches.\par

\indent Among inelastic interactions of $^7$Li nuclei in a track nuclear emulsion that were accelerated to a momentum of 3~GeV/$c$ per nucleon, events of the cluster fragmentation of $^7$Li nuclei into the $^3$H+$^4$He channel without meson production and without apparent excitation of target nuclei, this being indicative of a coherent fragmentation of projectile nuclei, were detected in \cite{Adamovich,Peresadko}. The vector sum of the transverse momenta of relativistic fragments that was measured experimentally in those events is the transverse-momentum transfer in the respective reaction, $\mathbf Q$. The measured dependence of the differential cross section for the reaction in question on the momentum transfer $Q$ was analyzed in \cite{Peresadko} under the assumption of the two-cluster structure ($^3$H+$^4$He) of the $^7$Li nucleus \cite{Kukulin,Dubovichenko} and on the basis of the formalism of Glauber-Sitenko theory for the case of involvement of nuclear clusters \cite{Davydovsky}. This analysis made it possible to reveal that this dependence of the differential cross sections with respect to $Q$ was different in form in inelastic and elastic diffraction processes and resulted in drawing conclusions on the role of electromagnetic and nuclear interactions and in pinpointing facets of interest for their further study, which included the Coulomb peak region standing out at small $Q$ and investigation of commensurate diffraction-cross-section oscillations predicted for inelastic processes on pure target nuclei. Experimental data on the fragmentation of relativistic nuclei on a proton target may serve as an important test of the theoretical model used to describe the two-cluster fragmentation of light nuclei, as well as a test of the method used to identify the channels under study. In the present article, which reports on our investigations pursuing further those described in \cite{Adamovich,Peresadko}, we give an account of the result of searches for and study of the reaction $^7$Li+$p\rightarrow^3$H+$^4$He+$p$ in track nuclear photoemulsion.\par

\section{\label{sec:level2}EXPERIMENTAL PROCEDURE}

\indent In a BR-2 track nuclear photoemulsion, there are $2.97\times10^{22}$~hydrogen atoms per 1~cm$^3$. In the track emulsion irradiated with $^7$Li relativistic nuclei, projectile-nucleus fragmentation on free protons must therefore also be observed in addition to nucleus-nucleus collisions. Processes involving the fragmentation of light nuclei on a proton target in a track nuclear photoemulsion are also being studied by the BECQUEREL Collaboration \cite{web}. A recoil proton must be observed in the interaction of a projectile nucleus with a proton. It is common practice to assume that, at the vertex of nucleus-nucleus interaction in a track emulsion, a recoil nucleus may be reliably detected if its range exceeds 2~$\mu$m. The momentum of a proton is about 20~MeV/$c$ if its range is 2~$\mu$m and about 30~MeV/$c$ if its range is 5~$\mu$m. Such a low energy threshold for proton detection makes it possible to study the fragmentation process in a track emulsion on a proton target over the entire momentum-transfer range, including the region of low momenta. For the sake of comparison, we note that, in studying the dissociation of a $^{12}$C nucleus to three alpha particles on a proton target in a propane bubble chamber \cite{Belaga}, recoil protons were detected only if their momenta exceeded 150~MeV/$c$. Investigation of an inelastic process on a proton target in a track emulsion is also of interest from the methodological point of view. In measuring the range in a track emulsion and in determining the momentum of the target nucleus in the final state, one knows the transverse momenta of all particles, including relativistic fragments and the target nucleus. The reconstruction of interaction kinematics makes it possible to test the reliability of fragmentation-channel identification and the validity of the procedure used to analyze processes involving the fragmentation of relativistic nuclei.\par 	

\indent In the present experiment, use was made of an emulsion chamber formed by layers of a BR-2 track nuclear photoemulsion, which is sensitive to the minimum ionization of singly charged particles. Emulsion layers had a thickness of about 600~$\mu$m, and their linear cross-sectional dimensions were 10$\times$20~cm$^2$. The emulsion chamber was exposed to a beam of $^7$Li nuclei accelerated to a momentum of 3~GeV/$c$ per nucleon at the synchrophasotron of the Joint Institute for Nuclear Research (JINR, Dubna). During irradiation, the track-emulsion layers were arranged to be parallel to the beam axis, so that beam particles entered the emulsion chamber from its endface and traversed the emulsion layers along their long sides. Tracks of $^7$Li relativistic nuclei could be traced with a microscope from the locus where $^7$Li nuclei entered a track-emulsion layer either to an event of nucleus-nucleus interaction or to their escape from the trackemulsion layer. The total length of all traced tracks, $L$, was used to determine the mean range of the nuclei and the reaction cross section. The charges of relativistic fragments of $^7$Li nuclei were determined visually by their track-ionization density, which differed by a factor of about four for singly and doubly charged particles. The direction of the relativisticnucleus track was determined from the coordinates of interaction vertex and the coordinates measured for several points on the track over a length of up to 2~mm from the interaction vertex. The direction of motion of a relativistic fragment was determined from the coordinates of the interaction vertex and the coordinates measured for the points on the track at distances of 500 and 1000~$\mu$m from the interaction vertex. For the results of the measurements, we took values averaged over several measurements. In some individual measurements, the scatter of the fragment polar angle $\theta$ about the primary direction of motion of the $^7$Li nucleus was about 0.03$^{\circ}$, while the scatter of values of the azimuthal angle $\psi$ in the plane orthogonal to the direction of motion of the $^7$Li nucleus was about 3$^\circ$. The mass of a relativistic fragment was determined by the method of measurement of multiple Coulomb scattering of the fragment in the track emulsion in the horizontal plane. In order to determine the average particle-scattering angle, the $Y$ coordinates of the track being considered were measured consecutively along the track at a distance $l$ between the points by using an MPE11 measuring microscope. The second differences of the $Y$ coordinates, $D$, characterize consecutive deflections of the track in the horizontal plane, while the ratio $D/l$ characterizes the scattering angle over the length $l$. In the case of the multiple Coulomb scattering of a charged particle, scattering angles and second differences of $Y$ coordinates obey a normal distribution, while the average value $\langle|D|\rangle$ over a cell of length $l$ has the form $\langle|D|\rangle=KZ_fl^{3/2}/(p\beta c)$, where $Z_f$, $p$, and $\beta c$ are the particle charge, momentum, and velocity, respectively. For a track emulsion of the BR-2 type, use is usually made of the experimentally determined scattering constant, which is equal to $K=28.5$. The values of $D$ and $l$ are given in $\mu$m units, while the value of $(p\beta c)$ is measured in GeV units. This relation between the measurable quantity $|D|$ and quantity $(p\beta c)$ makes it possible to estimate particle momenta and to perform a mass separation of fragments having the same charge. At the nucleon momentum of 3~GeV/$c$, the value of $\beta$ is close to unity, so that the distribution of particles with respect to $(p\beta c)$ reflects the momentum distribution almost completely. The experimental distribution of singly charged relativistic fragments with respect to $(p\beta c)$ can be satisfactorily approximated by the sum of three Gaussian functions that have maxima at values close to the momenta of single-, two-, and three-nucleon projectile fragments. According to the distribution, the value of 7~GeV was taken to be theminimumvalue of $(p\beta c)$ for $^3$H fragments. According to the distribution of doubly charged fragments with respect to $(p\beta c)$, the value of 10~GeV was taken to be the minimum value of $(p\beta c)$ for $^4$He fragments. The procedure used here to identify relativistic fragments was described in more detail elsewhere \cite{Adamovich}.\par

\begin{table}
\caption{\label{Table:1} Events of fragmentation of $^7$Li nuclei on a proton target}
\label{Table:1}       
\begin{tabular}{c|c|c|c|c|c|c|c|c|c}
\hline\noalign{\smallskip}
& \multicolumn{2}{c|}{$^7$Li$\rightarrow^3$H+$^4$He} & \multicolumn{5}{c|}{Recoil proton} & $Q-$ & Relative \\
Event& Moment- & Azimuthal & Range & Momentum & Azimuthal & Polar & Momentum & $P_t(p),$ & angle \\
&tum $Q$,&angle&$R(p)$,&$P(p)$, & angle $\psi_p$, & angle $\theta_p$, & $P_t(p)$, & MeV/$c$ & $\psi_Q-\psi_p$, \\
& MeV/$c$ & $\psi_Q,$~deg & $\mu$M & MeV/$c$ & deg & deg & MeV/$c$ &  & deg \\
\noalign{\smallskip}\hline\noalign{\smallskip}
65-041 & 109 & 349 & 303 & 115 & 181 & 82.6 & 114 & -5 & 180 \\
66-257 & 146 & 197 & 640 & 143 &  19 & 79.8 & 141 & +5 & 178 \\
69-381 & 241 & 278 &$>$4209&$>$247 & 104 & 79.9 &$>$243 &(-2)& 174 \\
70-373 & 244 &  75 &4600 & 252 & 249 & 79.1 & 247 & -3 &-174 \\
71-287 & 284 & 208 &$>$8181&$>$302 &  24 & 77.1 &$>$294 &(-10)&176 \\
71-371 & 208 & 329 &2958 & 224 & 153 & 80.9 & 221 &-13 & 176 \\
71-462 & 239 & 119 &$>$2487&$>$213 & 297 & 81.9 &$>$211 &(+25)&-178 \\
73-039 & 135 & 302 & 552 & 137 & 126 & 84.4 & 135 & 0 & 176 \\
73-144 & 322 & 339 &13825& 353 & 153 & 79.0 & 346 &-24 &-174 \\
74-066 & 185 & 111 &1602 & 187 & 285 & 81.3 & 185 & 0 &-174 \\
75-150 & 195 & 146 &2013 & 200 & 333 & 82.0 & 198 & -3 &173 \\
70-327 & 230 & 230 &5840 & 275 & 53  & 83.1 & 273 &-43 &177 \\
75-185 & 239 & 263 &5650 & 272 & 83  & 81.4 & 269 &-30 &180 \\
\noalign{\smallskip}\hline
Average& 214$\pm$5&&&227$\pm$6&&81$\pm$2&223$\pm$6&-8$\pm$2&176\\
\noalign{\smallskip}\hline
\end{tabular}
\end{table}

\section{\label{sec:level3}RESULTS OF THE MEASUREMENTS}

\indent In the fragmentation of a $^7$Li nucleus through the $^3$H+$^4$He channel on a proton target, there are no neutrons in the final state. It follows that, in the azimuthal plane, the total transverse momentum of the relativistic system and the transverse momentumof the target proton are equal in magnitude and have opposite directions. The equality of the absolute values of these vectors proves that the interaction of the relativistic nucleus being considered occurred in a collision with a free proton. In the cluster-fragmentation process, relativistic fragments preserve, to a precision of a percent, the absolute value of the primary projectilenucleus velocity before the interaction event. Therefore, the transverse momentum $p_t(A_f)$ of a fragment whose mass number is $A_f$ is given by $p_t(A_f)=p_0A_fsin\theta_f$, where $p_0$=3~GeV/$c$ is the momentum of a projectile nucleon and $\theta_f$ is the fragment polar emission angle with respect to the primary direction ofmotion of the relativistic nucleus before interaction. The accuracy in determining the fragment transverse momentum is determined by the accuracy in measuring the fragment polar angle and is about 7~MeV/$c$. According to estimations on the basis of a series of repeated measurements, the accuracy in determining the vector sum of the transverse momenta of relativistic fragments, $Q$, is about 10~MeV/$c$. The accuracy in determining the azimuthal angle of the total momentum of fragments, $\psi_Q$, is estimated at 3$^\circ$. The direction of the recoil track is determined on the basis of the coordinates measured by an ocular micrometer in themicroscope sight window of width about 30 $\mu$m. In a series of repeated measurements, the scatter of the values of the track polar angle $\theta_p$ does not exceed 1$^\circ$, the values of the azimuthal angle $\psi_p$ also being in agreement within 1$^\circ$.\par

\indent Among 3730 detected events involving the inelastic interaction of $^7$Li relativistic nuclei, we found 13 interaction events in which, in addition to $^3$H and $^4$He and relativistic fragments, there was a nonrelativisticparticle track whose direction and the vector sum of the transverse momenta of relativistic fragments, $\mathbf Q$, formed an angle of 180$^\circ$ in the azimuthal plane within the errors. The difference of the azimuthal angles of the vector $\mathbf Q$ and the vector $\mathbf P_t(p)$ of the recoil-proton transverse momentum $(\psi_Q-\psi_p)$ is presented in the rightmost column of the table for each event. The features of the system of relativistic fragments, the features of the recoil proton, and their comparison are also given in the table for each event. The values of the momentum $Q$ and of the azimuthal angle $\psi_Q$ are quoted for the system of relativistic fragments. For a nonrelativistic particle, we present the measured range $R(p)$ and the proton momentum $P(p)$ corresponding to this range, as well as the corresponding proton emission angles $\theta_p$ and $\psi_p$ and transverse momentum $P_t(p)=P(p)sin\theta_p$. In order to determine the recoil-proton momentum, we employed the dependence of the proton momentum on the proton range in the Ilford track emulsion. The accuracy in measuring ranges and in determining respective momenta was estimated at about 2\%.\par

\begin{figure}
    \includegraphics[width=4in]{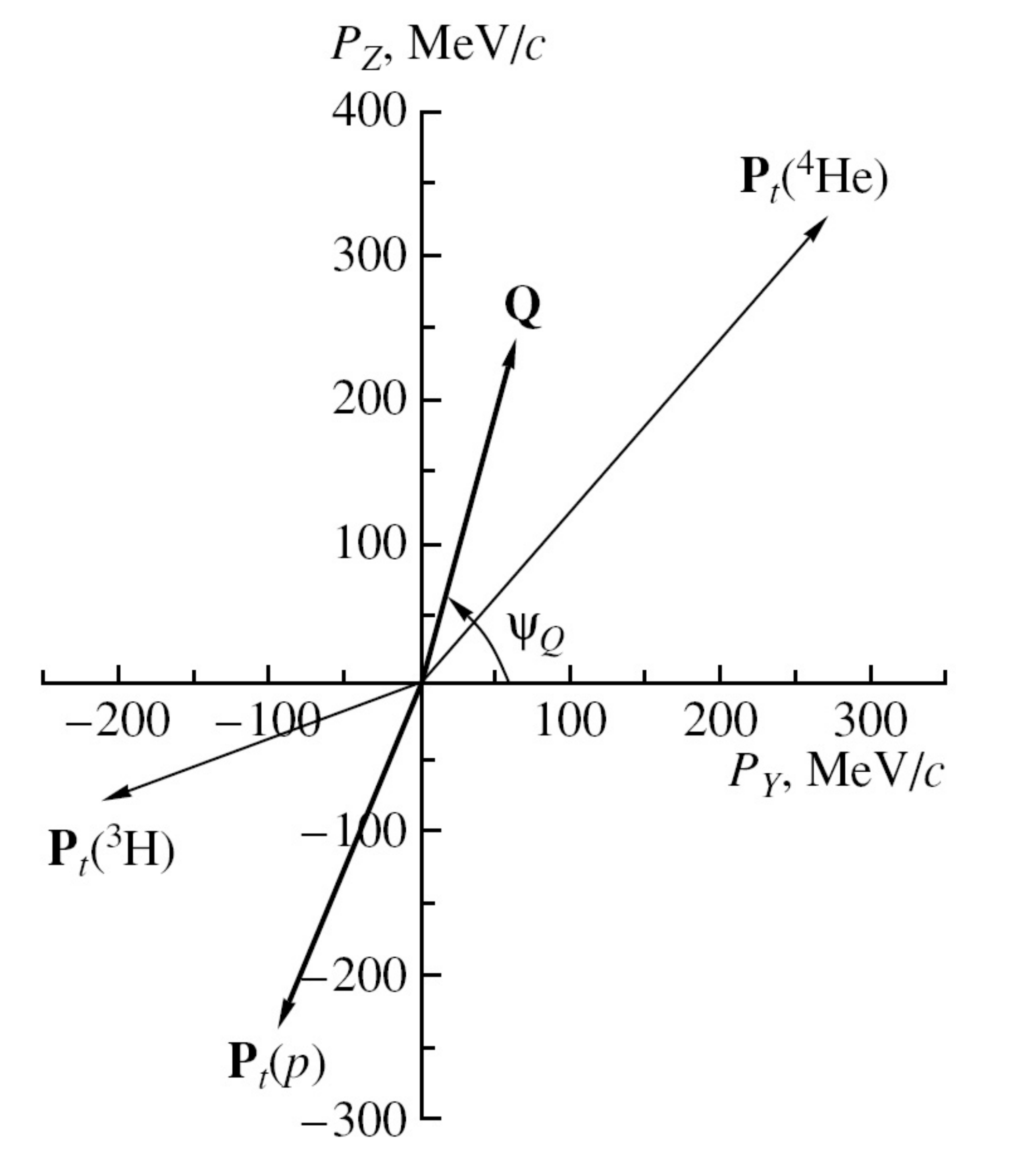}
    \caption{\label{Fig:1} Vectors of the transverse momenta of the $^3$H and $^4$He fragments, their vector sum $(\mathbf Q)$, and vector of recoilproton transverse momentum $[\mathbf P_t(p)]$ in the azimuthal plane in the 70-373 event.}
    \end{figure}
		
\indent The total transverse momentum of relativistic fragments is compared with the recoil-proton transverse momentum obtained under the assumption that the proton appears to be a nonrelativistic particle. The table gives the differences of the absolute values of the vector $\mathbf Q$ and the recoil-proton transverse momentum $P_t(p)$. By way of illustration, Fig. 1 shows the arrangement of the vectors of transverse momenta of relativistic fragments, the vector of their sum ($\mathbf Q$), and the vector of the recoil-proton transverse momentum $P_t(p)$ in the azimuthal plane in the 70-373 event. The azimuthal angles of the vectors are reckoned from the horizontal axis $Y$, which lies in the track-emulsion plane. Figure 1 shows the azimuthal angle $\psi_Q$ of the vector $\mathbf Q$. In ten events, the nonrelativistic-particle track terminates within the photoemulsion. In seven of these ten events, the difference of the absolute values of transverse momenta does not exceed 15~MeV/$c$. In the remaining three events, the proton momenta exceed the total momenta of relativistic fragments by 24, 30, and 43~MeV/$c$. In those three events, the proton ranges in the track emulsion exceed 5000 $\mu$m. This excess in the estimate of the proton momenta is possible if the braking power of the track emulsion used in the present experiment is lower than the braking power of the Ilford track emulsion. The inclusion of these three events increases significantly the variance of the difference of the momenta $Q$ and $P_t(p)$. The degree to which the transverse momenta of the relativistic system of fragments ($Q$) agree with the proton transverse momenta [$P_t(p)$] in these ten events can be further clarified in Fig. 2, where the events being considered are represented in a plane by points whose horizontal coordinates are values of $P_t(p)$ and whose vertical coordinates are values of $Q$. The line in Fig. 2 corresponds to the equality of the absolute values of the transverse momenta $Q$ and $P_t(p)$, and the magnitude of the deviation of a point from this line characterizes the difference of their values in an event. It can be seen that the points are concentrated in the vicinity of the line on which the momenta are equal. The observed agreement between the measured angular and momentum features of interactions on a proton target confirms correct identification of relativistic fragments and a satisfactory precision of angular measurement. The range of the nonrelativistic particle was not measured in three events. In two of these events, particles leave the track-emulsion stack, while, in one event, we were unable to trace the nonrelativistic track in the photoemulsion until the stop of the particle. In those cases, the minimum value of the momentum $P(p)$ was estimated on the basis of the length of the track to the point at which the particle escapes from the track emulsion. In those events, the difference of the transverse momenta $Q$ and $P_t(p)$ then proves to be -2, -10, and 25~MeV/$c$. In the table, these values are given parenthetically.\par

\begin{figure}
    \includegraphics[width=3in]{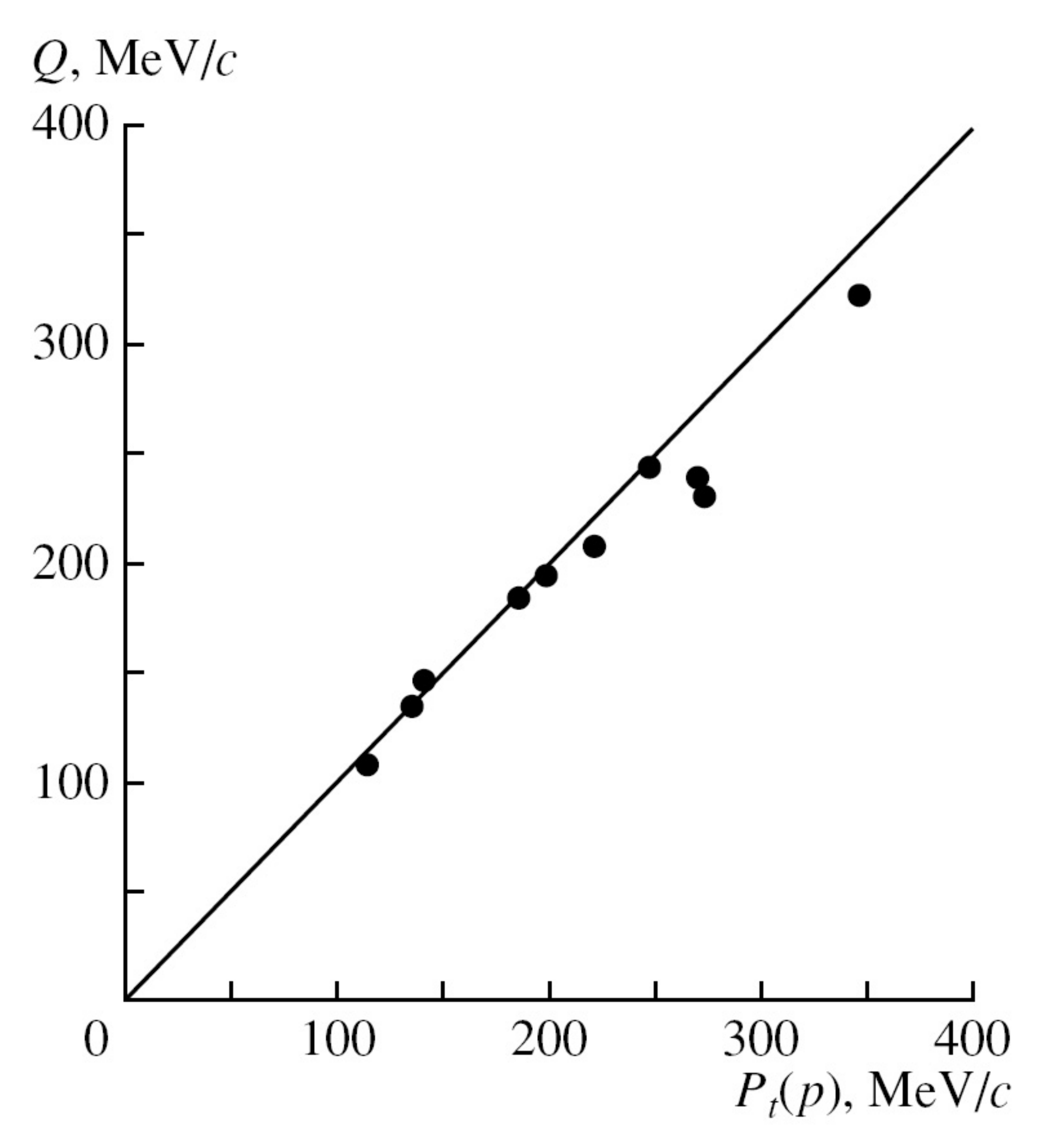}
    \caption{\label{Fig:2} Comparison of values of the recoil-proton transverse momentum $[P_t(p)]$ and the total transverse momentum of relativistic fragments $(Q)$. The displayed points represent events for which the horizontal coordinate is $P_t(p)$ and for which the value of $Q$ serves as the vertical coordinate. The straight line corresponds to the equality of the absolute values of the momenta $Q$ and $P_t(p)$.}
    \end{figure}
	
\indent The reaction cross section $\sigma(p)$ was determined as the ratio of the number of detected events, $N_{event}$, to the number of nuclei, $N_{nucl}$, in the track emulsion over the total length $L$ of the tracks traced in the experiment. The total length $L$ is 548~m. The proton density $\rho(p)$ in the track photoemulsion is $2.97\times10^{22}$~cm$^{-3}$. If we employ 13 events to estimate the cross section for the fragmentation of the $^7$Li nucleus through the $^3$H+$^4$He channel on a proton target, then the result for this cross section proves to be $\sigma(p)=N_{event}/(\rho(p)L)=8\pm2$~mb. The range for this reaction in the track emulsion is $\lambda(p)=42\pm12$~m. Using the cross section measured previously for nuclear fragmentation through this channel on track-emulsion nuclei ($29\pm3$~mb) and the powerlaw dependence of the cross section on the targetnucleusmass number in the form $\sigma(A)=\sigma(p)A^{\alpha}$, we find that the exponent is $\alpha=0.33\pm0.05$. The ratio of the experimental value of the cross section for the fragmentation of the $^7$Li nucleus through this channel on a proton target and its counterpart calculated theoretically within the two-cluster model of the $^7$Li nucleus may serve as an estimate of the probability for this clustering of nucleons in the $^7$Li nucleus. Experimental estimates of the probabilities for nucleon clustering in nuclei are performed on the basis of an analysis of cross sections for the quasielastic scattering of high-energy hadrons on the corresponding clusters in nuclei. For example, estimates of the numbers of quasideuterons and quasitritons in the $^{6,7}$Li nuclei were obtained in \cite{Abramov} on the basis of an analysis of the cross sections measured for the quasielastic scattering of $\pi^-$ mesons with momenta of about 1~GeV/$c$ on $^{6,7}$Li nuclei by using the 3-m magnetic spectrometer of the Institute of Theoretical and Experimental Physics (ITEP, Moscow), a scattered meson and a deuteron or a triton being detected in an event in these measurements.\par

\begin{figure}
    \includegraphics[width=3in]{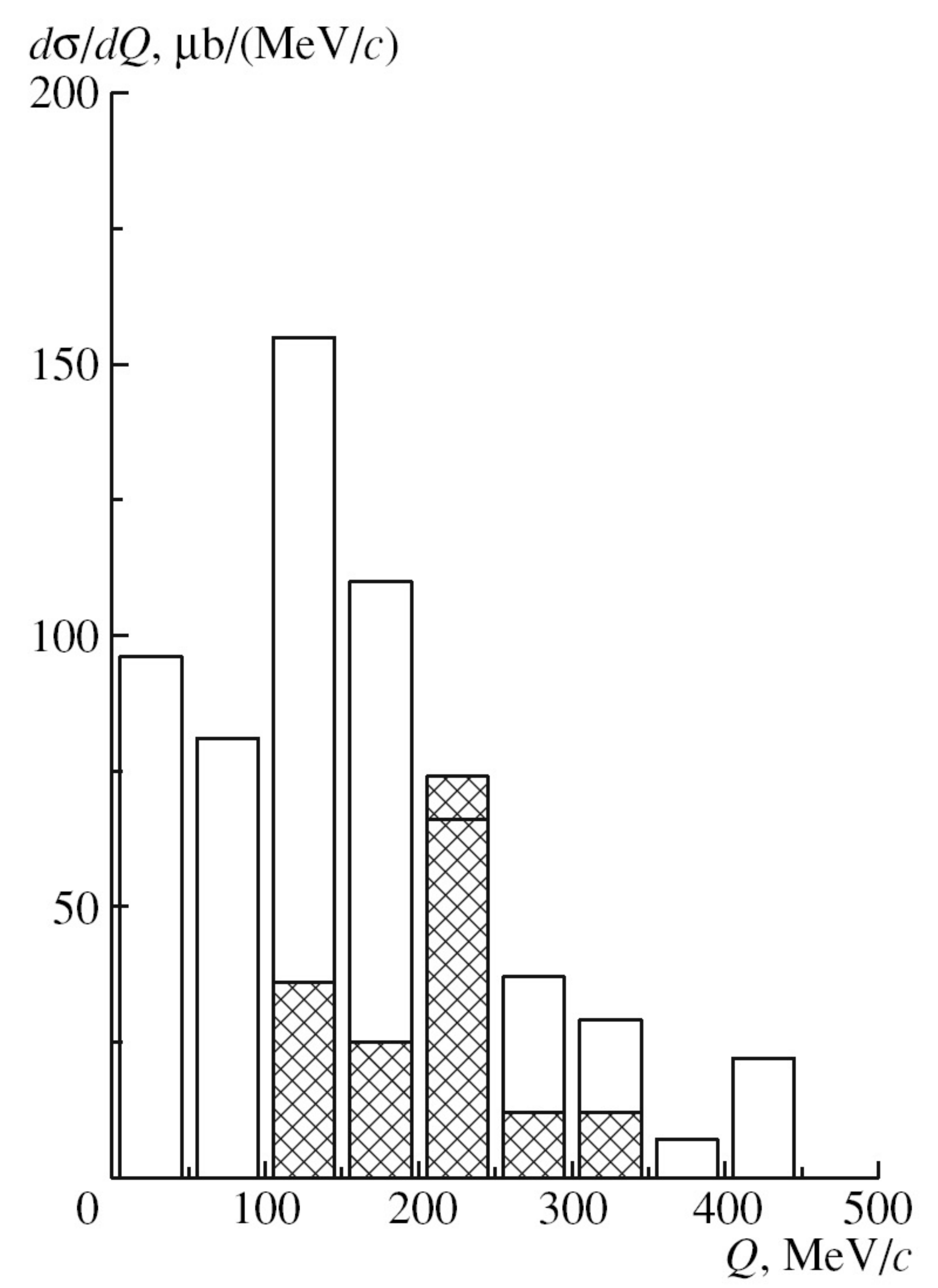}
    \caption{\label{Fig:3} Differential cross section for the fragmentation of $^7$Li nuclei through the $^3$H+$^4$He channel with respect to the transverse momentum $Q$. The shaded histogram represents the cross section in the case of a proton target, while the unshaded histogram corresponds to the cross section for the coherent fragmentation of $^7$Li nuclei on track-emulsion nuclei.}
    \end{figure}
	
\indent In Fig. 3, the shaded histogram represents the differential cross section with respect to the transversemomentum transfer $Q$ for the $^3$H+$^4$He channel of fragmentation on a proton target. In more than 80\% of events, the momenta are within the range from 100 to 250~MeV/$c$. The maximum momentum value does not exceed 400~MeV/$c$. The absence of events where the transverse momentum is below 100~MeV/$c$ is noteworthy; at the same time, more than onefourth of events fall within this region in coherent interactions on track-emulsion nuclei. The average value of the transverse-momentum transfer to relativistic fragments of $^7$Li nuclei in the case of a proton target is $\langle Q\rangle=214\pm5$~MeV/$c$, which exceeds markedly the average value of the transverse momentum for coherent interactions with track-emulsion nuclei ($166\pm5$~MeV/$c$). For the sake of comparison, the differential cross sections for the coherent fragmentation of $^7$Li nuclei through this channel on track-emulsion nuclei \cite{Peresadko} are also given in Fig. 3 (unshaded histogram).\par

\indent In the interaction of a relativistic nucleus with a free proton, one determines the total momentum of the recoil proton on the basis of the proton range. In such events, one can therefore find the ratio of the transverse- and longitudinal-momentum transfers in the reaction. These data make it possible to estimate the change in the fragment momenta because of the energy and momentum transfer from the relativistic projectile nucleus to the target nucleus. The proton emission angles $\theta_p$ in these events are distributed over the interval from 77$^\circ$ to 85$^\circ$. The average value of the polar proton emission angle $\langle\theta_p\rangle$ is 81$^\circ$. According to estimates based on 13 events, the average value of the total proton momentum is $227\pm6$~MeV/$c$, while the average value of the transverse momentum is $223\pm6$~MeV/$c$. The average value of of proton longitudinal momenta is $37\pm2$~MeV/$c$, while the variance of the longitudinal-momentum distribution is 16~MeV/$c$. These results indicate that the approximation where, in describing diffractive processes, one employs the reaction transverse momentum instead of the total momentum transfer would be legitimate here.\par

\section{\label{sec:level4}CONCLUSIONS}	

\indent In a nuclear track emulsion irradiated with $^7$Li nuclei accelerated to a momentum of 3~GeV/$c$ per nucleon, we have detected 13 events in which $^7$Li nuclei interacted with free protons, decaying to $^3$H and $^4$He fragments. The cross section for this channel of fragmentation on a proton target was found to be 8$\pm$2~mb. Note that the results concerning the fragmentation processes on a proton and on complex track-emulsion nuclei \cite{Peresadko} were obtained by using the same procedure in a single experiment.\par

\indent Taking into account the cross-section value of $29\pm3$~mb previously measured for nuclear fragmentation through this channel, we can describe the dependence of the cross section on the target-nucleus mass number by a power-law function whose exponent is $0.33\pm0.05$. The average value of the total transverse-momentum transfer to relativistic fragments is $214\pm5$~MeV/$c$. This value substantially exceeds the average value of the transverse momentumin coherent interactions with track-emulsion nuclei (the latter is $166\pm5$~MeV/$c$). The transverse component of the recoil-proton momentum is on average 98\% of the total momentum. The average value of proton longitudinal momenta is $37\pm2$~MeV/$c$, while the variance of the longitudinal-momentum distribution is 16~MeV/$c$. The experimental data obtained in the present study may serve as yet another test of correctness of the clustering approach in the theory of fragmentation at relativistic energies. The method used in the present study to identify the fragmentation of $^7$Li nuclei through the $^3$H+$^4$He channel can be applied in studying the cluster fragmentation of radioactive nuclei—in particular, in studying the fragmentation of $^7$Be nuclei through the $^3$He+$^4$He channel.\par

\begin{acknowledgments}
\indent This work was supported by the Russian Foundation for Basic Research (project no. 07-02-00871-à).\par
\end{acknowledgments}

\end{document}